\documentstyle[12pt,overcite,procsla]{article}

\def\etal{{\it et al.~}}

\def\ltsima{$\; \buildrel < \over \sim \;$}
\def\simlt{\lower.5ex\hbox{\ltsima}}
\def\gtsima{$\; \buildrel > \over \sim \;$}
\def\simgt{\lower.5ex\hbox{\gtsima}}

\def\chaphead{}
\newcount\eqnumber
\def\new{{\rm(\chaphead\the\eqnumber}\global\advance\eqnumber by 1}
\def\backref#1{\advance\eqnumber by -#1 (\chaphead\the\eqnumber
     \advance\eqnumber by #1 }
\def\last{\advance\eqnumber by -1 {\rm(\chaphead\the\eqnumber}\advance
     \eqnumber by 1}
\def\eq#1{\advance\eqnumber by -#1 equation (\chaphead\the\eqnumber
     \advance\eqnumber by #1}

\def\eqnam#1{\immediate\write1{\xdef\string#1{(\chaphead\the\eqnumber}}
   \xdef#1{(\chaphead\the\eqnumber}}

\begin{document}

\begin{Titlepage}

\Title
Application of Millisecond Pulsar Timing to the \\
Long-Term Stability of Clock Ensembles
\endTitle
\Author{Demetrios. N. Matsakis}
United States Naval Observatory \\
Washington, DC  20392-5420 \\
E-mail: dnm@orion.usno.navy.mil
\endAuthor
\And
\Author{Roger. S. Foster}
Remote Sensing Division, Code 7210 \\
Naval Research Laboratory\\
Washington, DC 20375-5351 \\
E-mail: foster@rira.nrl.navy.mil
\endAuthor
\typeout{ >> MG7tex note: note the hard space after the footnote here.}
\endAuthors

To be published in {\it Amazing Light}, by Springer-Verlag Press, 1995.
\vskip 12pt

\begin{Abstract}
Terrestrial timescales show instabilities due to the
physical limitations of the atomic clocks.  Stricter environmental isolation
and increased numbers of improved cesium clocks and cavity-tuned hydrogen
masers
have resulted in time scales
more accurate by a factor of about five.  The use of different clock
ensembles results in measurable changes in some millisecond pulsar
timing data.  We investigate the possible
application of millisecond pulsars to define a precise
long-term time standard and positional reference system in a nearly
inertial reference frame.   Although possible quantitative contribution of
the two longest studied millisecond pulsars to terrestrial timescales
appears minimal, they may prove useful as independent standards in
identifying error sources that are difficult to detect due to the finite
lifetime and common reference frame of terrestrial clocks.
New millisecond pulsars, perhaps some with even better timing
properties, may be discovered as a result of the current global
pulsar search efforts.
\end{Abstract}
\end{Titlepage}

\section{Introduction}

The 1982 discovery of the 1.6 millisecond pulsar
B1937+21\cite{b82} has
provided an object with rotational stability comparable to the best
atomic time standards over periods exceeding a few years\cite{ktr94}.
The microwave beams of millisecond pulsars, which are rigidly anchored to
rotating neutron stars, can act as precision celestial
clocks owing to the combined effects of their large rotational
energies, $10^{50-52}$ erg, and low energy loss rates.
Millisecond pulsars provide an
opportunity for astronomical observations to define International
Atomic Time (TAI) over long periods.

Because each pulsar has a different spin and spin-down rate, timing
measurements of millisecond pulsars cannot provide a more
{\it accurate} timescale than current atomic timescales.  Since the best
foreseeable daily-averaged pulsar timing measurements will be accurate to 0.1
microsecond, two observations separated by $5\times 10^7$ seconds would be
needed to reach the $\rm 2.5\times10^{-15} s/s$ TAI monthly {\it precision}
\cite{t94} (defined as the internal rms of the data set,
ignoring systematic errors), although more frequent pulsar observations would
reduce this considerably.

The existence of a number of millisecond pulsars distributed across the sky
leads to the possibility of timing several pulsars against each other,
with terrestrial clocks providing merely the endpoint times and a means
of interpolating observations\cite{fb90, gp91, ptt93}.
Such a program can smooth out variations in TAI, better determine the
masses of the outer planets\cite{m71}, place
limits on perturbations produced by a primordial spectrum of
gravitational radiation\cite{b95, srtr90},
constrain the nature of
interstellar turbulence\cite{fc90}, and provide an
astrometric tie for the planetary
and radio reference frames\cite{lrwn90}.

The search for new millisecond
pulsars is a matter of intense activity and steady progress.
More than 30 millisecond pulsars have been found in globular clusters, and
over 25 have been found in the galactic disk, over a wide range of
galactic latitudes and longitudes.  Unlike slow pulsars, more than 75\% of
the millisecond pulsars have stellar companions; millisecond pulsars
are often termed recycled pulsars because it
is believed they were spun-up by their companions. \cite{gl92} It is estimated
that over 30,000 in the galaxy are beaming in our
direction\cite{bl95}.
Non-cluster millisecond pulsars are discovered at a rate of about
1 per 200-300 square degrees\cite{fcwa95}.  Assuming
a factor of two improvement in search
efficiency and ignoring serendipity, perhaps up to 400
field millisecond pulsars could be discovered within the next few decades.

In this work we summarize the problems and status of terrestrial timescales
and pulsar observations, and explore the contributions pulsars can
make to terrestrial timescales.

\section{Terrestrial Clocks}

According to the International System (SI), the second is defined
so that the frequency of
the $(F,m_f)$=(3,0) to (4,0) hyperfine transition of cesium 133
is exactly 9.192631770
GHz, in the absence of a magnetic field and located on
the geoid\cite{tgcwm67}.
As recognized in the definition, the great majority of the
timing data used to compute TAI are from cesium-based frequency
standards\cite{r83}.  Cesium atoms in the (3,0) state are
created in an oven and
isolated from cesium atoms in other states by their deflection
in an inhomogeneous
magnetic field.  They are then allowed to travel through
a magnetically shielded
cavity, in which they are exposed to 9 GHz microwaves at each end.
The probability that the 9 GHz microwaves will induce a
transition to the (4,0) state has a Ramsey-pattern dependence upon frequency
of the microwaves\cite{r63};
the number of ions which have made the
transition to the (4,0) state is measured by
the current created after deflection by an inhomogeneous field
at the exit point of the cavity.
The principal error sources are the magnetic field correction
and the phase difference
between the microwave fields at the two interrogation
regions\cite{bauch87a}.
Recently, Hewlett-Packard
introduced their model 5071 cesium standards, with a significantly improved
electronics package\cite{cg92}.
This model cesium clock has demonstrated superior performance both in
the laboratory\cite{br95} and in the field\cite{w94}.
The world's most accurate laboratory standards are maintained
by the Physikalisch-Technische Bundesanstalt (PTB) in
Germany and the National Institute of Standards
and Technology (NIST), which achieve long-term stability through use
of special beam-path and field switching techniques and the use of
a longer cesium travel path\cite{bauch87b}.

A second frequency standard used for high precision time keeping
is the hydrogen maser.
Hydrogen masers are based upon the 21-cm transition of atomic
hydrogen and can achieve much
greater precision in the short-term\cite{t51}.  However,
their long-term accuracy is limited by
variations in the dimensions of the
confining cavity and interactions with the cavity walls\cite{p73}.
Special Teflon coatings have been developed which minimize the
wall-shift.  Recently the Sigma Tau Standards Corporation has begun
marketing masers whose cavity capacitance is stabilized and optimized
through an attached varactor diode and a tuning circuit which
measures the maser amplitude at 7.5 kHz on each side
of the microwave resonance\cite{okmp92}.

     The precision of the best types of frequency standards at the U.S.
Naval Observatory (USNO)
as a function of the averaging time are shown in Figure 1,
which was adapted from Breakiron\cite{br95}.
Note that masers have significantly better performance on short timescales,
but on the longest scale, 60 days, their performance is similar to
the cesiums.
Following the methodology of Barnes \etal\cite{barnes71}, we summarize in
Table 1 the error performance
of the clocks as a sum of white and random walk frequency components.
Since these clocks have not been in existence long enough to gather
data on still longer timescales, it is likely that future analyses
will identify other components to the variation of the cesium standards.
A long-term analysis of maser data will be more difficult to perform since
they undergo discontinuous changes in frequency and frequency drift (the time
derivative of the frequency) on timescales of a few months, which are removed
by comparison with other clocks.

Within the next decade, we can expect
further improvements in terrestrial standards.  The closest to implementation
are optically pumped mercury stored-ion devices.
Their advantage is based upon the
high frequency of the
observed transition (40.5 GHz) and coherence times which
allow line-widths as low as 17 mHz.
Although the accuracy of prototypes was limited by vacuum
contamination\cite{mkdgc95}, recent designs based upon
a linear ion trap
have achieved frequency stabilities better than $\rm 10^{-15}s/s$ on
timescales of a few hours\cite{tpdm94} (frequency is defined hereafter
as the time derivative of the timing error in units of time, not phase).
Somewhat further away
are clocks based upon trapped yttrium\cite{fslcmb94}, single trapped mercury
ions\cite{whw91}, and cesium atomic fountains, which achieve long
coherence times by first cooling atoms in ``optical molasses''
to 700 nanokelvins and
then allowing them to rise out of the measurement region until gravity
brings them back\cite{kprsj95}.

\section{Global Time Transfer}

    The Global Positioning System (GPS) has within the last few years
become the chief instrument for comparing clocks
of different institutions, as well as for determining position on the Earth.
It consists of a constellation of 24 satellites, run by the
U.S. Air Force Second Satellite Operations Squadron, each containing
at least one cesium frequency standard
and circling the Earth twice per sidereal day\cite{d91}.  These satellites
constantly broadcast their coded position and timing information, in both
classified and unclassified channels,
enabling anyone with the proper equipment to determine, by
triangulation, the position and time at the receiver.  Depending
upon the kind of GPS receiver, the time can be communicated through
an output signal and/or compared digitally to the time input from a standard
at the site.  The individual timing characteristics of the
satellites are monitored by the USNO, which forwards the
information to the Air Force for use in determining the frequency
and time offset corrections and in measuring orbital
variations.  Although the specified GPS time precision is 1
microsecond, since January 1993 GPS time has
kept within 260 nanoseconds of the USNO Master Clock \#2 (UTC(USNO)),
and the rms difference over that period was
60 nanoseconds.  Satellites also broadcast on-line predictions
of GPS$-$UTC(USNO),
which are specified to be less than 90 nanoseconds, but are usually
below 20 nanoseconds (which
is close to the measurement error).  Evaluations of GPS$-$UTC(USNO)
are published as USNO Series 4, and are also
available via modem, Internet, and the World Wide Web home
page {\tt http://tycho.usno.navy.mil}.

     While the received GPS signal is distorted and delayed by unmodeled
tropospheric, ionospheric, relativistic, and multi-path effects,
the largest error contribution to unclassified
use is the intentional distortion of up to
340 nanoseconds imposed by the military for security reasons and
termed ``selective availability'' (SA).  SA consists of both the
insertion of a variable time delay and a misrepresentation of
satellite orbital parameters, which are used to determine the
propagation time for the signal to reach the observer.

     If one is merely interested in time difference between
remote clocks, it is possible to eliminate the error due to the
variable time delay through the ``Common View'' method.  In Common
View, observations are scheduled so that both sites observe the
same satellite at the same time.  While the difference neatly
cancels satellite timing errors, it cannot eliminate the errors
due to the incorrectly broadcast orbital parameters.  It is
possible to correct for this after the fact, using satellite
orbital elements routinely disseminated by the International GPS
Service for Geodynamics, which relies upon multiple observations
for its analysis.  Common View techniques routinely
provide time-transfer precision
of 5 to 15 nanoseconds\cite{k83, lt91}.

     While it is possible, in principle, to correct any
GPS measurement in real time for SA using
Common View, such scheduling
tends to result in fewer observations which are also conducted at lower
elevation angles, so that tropospheric and ionospheric modeling
errors are more significant.  At the USNO, unclassified GPS
transfers are achieved using the ``melting-pot'' method, which
forsakes Common View in order to obtain a straight forward average
of as many satellites as possible.  Even in the
presence of SA, a precision of 20 nanoseconds
is attainable if one is able to average at least seventy
13-minute satellite passes over a
48-hour period\cite{c93}.  The advantage
of this method is even greater for receivers which can observe more than
one satellite at the same time.  Another
form of time-transfer consists of using available non-GPS
satellites to send signals to and from the two remote sites.  Aside
from achieving sub-nanosecond time transfer precision, such studies often
reveal systematic and perhaps seasonal
differences with conventional GPS measurements,
of order 10 nanoseconds\cite{dey94}.

\section{Terrestrial Timescales and the BIPM}

   The goal of timescale formulation is to approximate
``true'' coordinate time, from an average of differential clock
timing measurements.  In essence, the clocks serve as frequency
standards to create a heterodyned tone of known or
defined value.  Individual timing events are realized either
through summation of cycles or use of  ``zero-crossings'' of the
sinusoidal output voltage.  In practice, timescales are usually generated
in the frequency domain, in which a mean
frequency is computed and used to define
the time from an arbitrary offset.

     Internationally, the official responsibility for timescale
generation lies with the International Bureau of
Weights and Measures (BIPM) in Paris, France as first established by
the Convention du Metre in 1875.
The BIPM produces TAI by independently evaluating
the individual characteristics of approximately
200 clocks maintained by almost 50 institutions\cite{bipm93}.
Evaluations are made at 10-day intervals,
and reported monthly in their Circular T.  The BIPM also
evaluates timescales generated by
many national institutions, including GPS time.
Some of these timescales are free-running,
such as the USNO ``A.1'' series, which is currently comparable
to TAI in accuracy.
Others, such as UTC(NIST) and
UTC(USNO), are steered so as to track UTC closely.
The UTC(USNO), which is the national standard for
legal and military matters, is available on-line and
is in practice based upon the output of a single
maser that is steered daily
through the addition of small frequency offsets
so as to approach the computed mean of
all contributing USNO clocks, which itself is adjusted
towards extrapolations of the BIPM-computed TAI$-$UTC(USNO). As a result
of this steering TAI and UTC(USNO) have since 1991 kept within 200 nanoseconds.

     Once all the time comparisons have been assembled, the
problem of timescale generation reduces to
calculating weights and offsets for each
standard.  The BIPM first generates a free-running timescale, the Echelle
Atomique Libre (EAL).  The EAL is computed by evaluating clock
weights each 60 days.  The weights
are determined from the variance of the frequency offsets over the past
year, after allowing for any intentional rate adjustments.
A maximum weight is
chosen so that the timescale will not depend upon just a few of
the clocks.  While both the general and specific procedures are periodically
under review, in 1994 about half of all clocks used had this maximum
weight, while most of the rest were weighted
less than 20\% of the maximum.  The BIPM does not subtract
offsets from the clock frequencies, and
this can lead to minor discontinuities
as clocks are added, subtracted, or reweighted.  The
TAI timescale is created by gently steering
the EAL so that its
long-term behavior conforms to the ``primary'' frequency standards, which
are currently the two ``long-tube'' primary
standards of the PTB\cite{bauch87b} and NIST-7\cite{bipm93}.  This
is a practical way for a real time system to take into
account the fact that the primary standards have greater long-term accuracy
than accounted for by the precision-based weighting system.
Again, there is a minor problem in that no allowance is made
for fact that the NIST-7 frequency is corrected
by about $\rm 1.9\times 10^{-14}s/s$ for the
AC Stark effect due to the microwave background
experienced by the cesium atoms\cite{dlgs93}, whereas
the PTB frequencies are not.

At the USNO, the A.1 timescale is currently determined using a
weighting function which weights clocks by their
type.  Recent maser data are given a much higher weight than
recent cesium data, but older maser data are progressively down-weighted,
with zero weight assigned
to maser data more than 60 days old\cite{br91}.
Also, data from the older model cesium standards are
included at 0.65 the weight of model 5071 cesium clocks.
Unlike the BIPM, each clock is introduced to the mean after subtracting an
initial frequency offset.  This has the effect of reducing discontinuities,
and the drawback of subtly over-weighting initial clocks is soon lost in the
random-walk noise as long as no systematic errors are present.
For historical reasons, the A.1 is intentionally steered
a constant 19 ns/day\cite{whp70}.

Using data from 1993 and 1994, the BIPM estimates the 100-day
error of the EAL as slightly less than the TAI, and both to be about
$\rm 2.5\times 10^{-15}s/s$\cite{t94}; the errors in the 1980's were
above $\rm 10^{-14}s/s$.  The recent improvement in both
TAI and the USNO free-running
timescale A.1 is evident in Figures 2-4, which gives their difference
as a function of time.  Part of this apparent improvement is due to the fact
that the contribution of the USNO clocks to
TAI has varied (from 30\% in the 1980's to
a low near 15\% in 1990 to almost 40\% in 1994) but
most is due to the world-wide increase in number, quality, and environmental
isolation of the individual frequency standards.
Similar increases in precision are shown in the
comparisons with the independent PTB timescale.

For practical reasons, other timescales are often used which differ
from TAI by specific and calculable amounts.
Coordinated universal time (UTC) differs from TAI only
by an integral number of leap seconds, which
crudely take into account variations in the rate of
Earth rotation\cite{e93}.  These leap seconds are extra seconds
included in UTC at UT midnight on
either December 31 or June 30 so as to minimize
UT1$-$UTC, where UT1 is time related to the rotational angle
of the earth after allowing for polar motion\cite{e93, m72, a82}.
They have lately been inserted at a rate approaching once a year, and on
July 1, 1994, TAI$-$UTC= 29 seconds.  Measurements and predictions of UT1,
polar motion, and nutation parameters \cite{neos94} are
available on-line from the home page {\tt http://maia.usno.navy.mil}.
In order to allow for continuity with ephemeris time (ET), Terrestrial
Time (TT, also called Terrestrial Dynamical Time, TDT) is defined
to be offset by exactly 32.184 seconds from the
``ideal'' form of TAI, which is what TAI
would be if no instrumental noise were present.
Although in the future TT may be realized from an average of many sources,
including pulsars, currently TT is realized only from TAI, by post-processing
the raw clock data and improving upon the initial weighting and steering
corrections\cite{g87}.  Since 1982, the BIPM-computed numerical
difference TT$-$TAI
has varied by over 3 microseconds absolutely,
and over 2 microseconds after
removing the quadratic component of the variation.  In order to
provide continuity in the interpretation of old data and allow for
the effects of general relativity in a consistent manner, the IAU
in 1991 passed resolutions recognizing timescales
which distinguish between centers of reference at the solar
system barycenter (Barycentric Coordinate Time, TCB), the
center of the Earth (Geocentric Coordinate Time, TCG), and the geoid
(TT, TAI, UTC, ET)\cite{iau91, sf92, f95}.

\section{The Techniques of Pulsar Timing}

     The precision of an individual pulsar time-of-arrival (TOA) measurement
depends upon its pulse period and the signal-to-noise ratio
as determined from the particular instrumental setups\cite{b94,
b91}.  The fastest pulsar known today is still PSR B1937+21,
whose 1.6 ms period is close to the theoretical limit determined
from balancing the centripetal
and gravitational forces, assuming
a typical neutron star equation of state at densities of
$\rm \simgt 10^{14}~g~cm^{-3}$\cite{fip86}.

The simplest rotation model for a solitary
neutron star can be given in terms of
phase residuals as a power series in time:

$$ \phi (t) = \phi_0 + \Omega (t-t_0) + \dot \Omega (t-t_0)^2/2 +
\ddot \Omega (t-t_0)^3/6 + ... \eqno\new )$$

\noindent
where $\phi$ is the pulse phase, $\Omega$ is the rotation frequency,
$\dot \Omega$ is the spin-down rate.  The spin frequency is related to
the pulse period by $\Omega = 2 \pi /P$ $(\Omega = 2 \pi \nu)$, while
$\dot \Omega = -2 \pi \dot P / P^2$.
If the spin-down of the pulsar
is driven by magnetic dipole radiation, due to misalignment of the
magnetic and rotational axes, then the braking index $n$, defined by
$  { \Omega \ddot \Omega / {\dot \Omega ^2} }
\equiv 2 - {  P \ddot P / { \dot P ^2} } $, is 3.
If the initial pulsar period was $P_0$, the pulsar age can be computed from

$$ \tau = { {P} \over {(n-1) \dot P}}
 [ 1 - ({P_0 \over P})^{(n-1)}].
 \eqno\new )  $$

\noindent
Most pulsars $\ddot P$ measurements yield only upper limits, but in those
pulsars for which $\ddot P$ can be measured above the noise,
$n$ is found to be between 2 and 3.  Inverting the equations
above, we derive a theoretical value

$$ \ddot \Omega = { n \dot\Omega ^2 \over \Omega }
= { 2n \pi \dot P^2 \over P^3 }.  \eqno\new )$$

\noindent
In the case of PSR B1937+21 the theoretical slowdown
would result in a 0.1 microsecond
residual from a constant $\dot P$ model
after 11 years.
The relativistic time-of-flight effects and acceleration due to the
galactic gravitational field can also make
small, but non-negligible contributions to  $\dot \Omega$ and
the theoretical $\ddot \Omega$ \cite{ctktr94}.
Also, for pulsars in globular clusters, contributions to
$\ddot \Omega$ from Doppler acceleration
can be important where the gravitational
potential of the cluster accelerates the pulsar,
influencing the rotation period over several-year timescales.
\cite{bra87, wol89}

Pulsar timing requires an explicit definition of both a steady atomic
time standard and accurate monitoring of the motion of the Earth with
respect to the background stars.  This task is accomplished by using a
clock whose offset is frequently calibrated against a real time
steered UTC service, such as UTC(NIST) or UTC(USNO).  This
real time UTC realization is later corrected to a more accurate
timescale using after-the-fact determinations published by the BIPM.
Although TAI is often used, TT is recommended as the most
accurate terrestrial scale available today\cite{gp91}.
Once a terrestrial timescale has been chosen, it must be converted to
the barycentric frame (TCB).  The omission of this time conversion
would, among other things, result in a sinusoidal monthly error of
about a  microsecond as the Moon pulls the
the Earth back and forth through the solar gravitational potential well.
In most software, the time conversion happens automatically as part
of the four-dimensional general-relativistic transformation which
accounts for the Earth's orbit.  The IAU has recommended that all users
avoid transformations to Barycentric Dynamical Time (TDB), \cite{iau91}
which is scaled to remove secular differences
with TT; as a result it diverges from TCB at a constant rate of
$1.55\times 10^{-8}$, which would be absorbed into the fitted parameters.

The geometric, general relativistic,
and astrometric effects due to the position and velocity of
the antenna with respect to
the solar system barycenter must also be removed.  In most cases
this is done using the
IERS values for UT1$-$UTC and the DE200 ephemeris\cite{s90}.
Since most observations used for computing the Earth's orbit
are Earth-based they are relatively insensitive
to the uncertainties in the masses of the outer planets.  The
most uncertain long-term component of the ephemeris for pulsar work
is the motion of the Sun and known planets about
the solar system barycenter.  The largest error contribution is
due to the uncertainty in the
mass of Saturn, which could result in a TOA error of order 7
microseconds with a 30-year period\cite{b94}.
This and other such errors would
be identifiable through their different and systematic projections upon the
pulsars observed\cite{fb90}.  Fukushima has examined
the differences between the DE200, DE102, and DE245 ephemerides and found
the differences between them (including their effects on the relativistic
time correction TCB$-$TT) to be negligible for pulsar work\cite{f95}.

Pulsar signals must be corrected for dispersion caused by
the intervening plasma using the frequency dependence
of the delay.  For a homogeneous and isotropic medium the group velocity
is given by $ v_g = c (1 - \omega_p^2/\omega^2)^{1/2}$,
where $\omega_p$ is the plasma frequency and $\omega$ is the wave
frequency.  The plasma frequency in Gaussian units is
$\omega_p^2 = 4 \pi n_e e^2/m, $
where $n_e$ is the mean electron density, and $e$ and $m$
are the electron charge and mass.
The delay is proportional to the column density of electrons in the
line of sight towards the pulsar.  This column density is called the
dispersion measure and is defined as
$ DM = \int^z_0 n_e dl,$
where $z$ is the pulsar distance.  The dispersion measure is a directly
measurable quantity determined from the differential pulse arrival time
between two frequencies.  To first order,
observing $v_g$ gives a time delay between two frequencies as

$$ t_2 -t_1 = { {2\pi e^2} \over {mc} } (\omega^{-2}_2 - \omega^{-2}_1)
DM. \eqno\new )$$

\noindent
Observations of dispersion measures toward 706 radio pulsars
give values that range from 1.8 to $\rm \sim 1000~cm^{-3}~pc$\cite{c95}.
For some lines of sight the turbulent electron population may
cause different wavefronts to travel along slightly different paths,
which can lead to delays that are not removable by the two-frequency formula,
although some improvement can occur with multi-frequency observations.
The unmodeled delay, which depends upon the total electron column density and
the turbulent properties on the interstellar electron population,
has been estimated in the worst cases to be
at the sub-microsecond level for radio frequency observations
at 1.4 GHz and above\cite{fc90}.

Another important consideration is that millisecond pulsars have steep
radio spectra and are generally weak enough that observations at the largest
telescopes are required in order to achieve an adequate signal-to-noise
ratio.  An estimate of the SNR achievable at current and future sites
is provided in Table 2.

Once the TOAs from a pulsar are
assembled and iteratively corrected for known effects, several pulsar
parameters are typically solved for.  Uncertainty in the {\it a priori} pulsar
period and spin-down rate are identified through their quadratic dependence.
The pulsar parallax is identified from the TOA by its semiannual periodicity,
position errors through their annual periodicity, and proper motion
by a growing annual periodicity.  Pulsar parameters associated with
orbits about companions are more complex\cite{dd86, wol94}.
More exotic error sources, such as gravitational
radiation and unknown planets, can be identified through comparison of
delays with other members of the pulsar ensemble.  If the TOA errors
are modeled as a multi-pole expansion over the sky, timing errors
are related to the monopole term,
ephemeris errors result in a dipole dependence, and gravitational
radiation would cause a quadrupole and higher order effect\cite{fb90}.

\section{Stability of Pulsar Time Standards}

Using data on the millisecond pulsars B1937+21 and B1855+09 made available
by anonymous FTP from Princeton University\cite{ktr94}, the program
TEMPO\cite{tw89} was used to solve for
the pulsar periods, spin-down rates, position, and proper motion.
The residuals to the solutions are shown in Figure 5.
As Kaspi \etal\cite{ktr94} found with the same data,
the residuals of PSR B1937+21
appear to display red noise (long-period correlations, in this case
well-modeled as a non-zero $\ddot P$), whereas the residuals
of PSR B1855+09 do not.  Since these two pulsars are close together in
the sky, the red noise is considered to be intrinsic to the pulsar
B1937+21 and not due to the timescale or ephemeris errors.
The timing noise for millisecond pulsars
seems to follow the correlation
with $\dot P$ found for slower pulsars by Arzoumanian \etal\cite{ktr94,
cblb94, antt94}:

$$ \Delta_8 = 6.6 + 0.6 {\rm log}\dot P , \eqno\new )$$

\noindent
where the ``stability parameter'' $\Delta_8 = {\rm log} \left( { 1\over {6\nu}}
|\ddot \nu| (10^8 {\rm s})^3\right)$ with a scatter of $\pm$ one decade.
Pulsar B1937+21 has a $\Delta_8$ = $-$5.5.  If this inverse correlation is
valid for millisecond pulsars, then the PSR J1713$+$0747 would be
a particularly promising candidate to have a very low $\Delta_8$ stability
parameter\cite{fcw95}, with a red noise less
than that of PSR B1937+21 by a factor of about 5.

The effects of different input timescales are shown in Figure 6, which plots
the difference between the residuals of parameter fits to PSR B1937+21,
which differ only in the input timescale assumed.  For these and subsequent
solutions, the parameters were the pulsar period, period derivative, position,
parallax, and proper motion.  In Figure 7 these
difference solutions are double-differenced with the differences between
the timescales, as made available by the BIPM, with a second-order
term removed.  The double-differences indicate how much timescale
variation is soaked up into the fitted parameters,
and hence is indeterminable from pulsar TOA data.  The annual
signature evident in the plots, for example, shows how pulsar
timing data alone will never be able to identify strictly periodic
annual variations in terrestrial standards.  In the presence of an
independent means of determining pulsar positions, proper motion,
or parallax (such as astrometric Very Long Baseline Interferometry, or VLBI),
such variations would be
recognizable as displacements in the fitted parameters.
Table 3 shows the current positional precision of a number
of millisecond pulsars, derived from timing data.
The coupling of the dynamic and radio reference frames will,
along with VLBI observations of millisecond pulsars, aid
in removing two degrees of freedom from the pulsar fitting models
and improve our ability to constrain other sources of errors.

Slightly extending the analysis of Blandford \etal\cite{bnr84}, it is possible
to apply a Weiner filter to the residuals by Fourier-transforming
the residuals, and then transforming back to the time domain after
multiplication by a ``transfer function,'' which corrects for
the distortion caused by fitting to pulsar parameters.
The exact correction factor would also include a factor incorporating
the clock and pulsar stabilities as a function of frequency.  Since pulsar data
are much less accurate on short time scales, compared to terrestrial standards,
pulsar data would be given insignificant weights at short periods.
For the generation
of the plots in Figure 8, a simpler analysis was used in which all periods
less than 400 days were discarded.  Figure 9 shows the ``transfer function''
for two different time ranges,
which indicate what frequencies in the
raw data are absorbed by the fitted parameters\cite{bnr84}.
The low values at periods of one year and six months indicate that
any noise in the data with those frequencies will be absorbed into the
fitted pulsar parameters.  It is evident that the accumulation of another
decade of data will bring about a considerable improvement in the ability
of pulsar timing data to discern long-term variations.

The contribution millisecond pulsar data can make to a timescale is
entirely dependent upon the quantity and quality of
the pulsar TOAs compared to the
terrestrial standards.  Included along with the simple clock model
in Table 1 are parameters for the timing stability of PSR B1937+21,
which is characterizable as a random walk in
frequency, or perhaps an integrated random
walk, which is even redder\cite{k94}.  These numbers were chosen
to be consistent with the residuals plotted in Figure 5, and they
show that PSR B1937+21 is less accurate than a
single model 5071 cesium standard.
If the relation between timing stability and $\dot P$ is
valid, then the pulsar J1713+0747 would be five times more accurate,
and an ensemble consisting of several such pulsars could approach the
accuracy of today's ensemble of terrestrial frequency standards.
More important than the numerical evaluations is the fact that
pulsars can provide an independent means of evaluating clock performance,
and can serve to constrain the amount of extremely red noise in terrestrial
clocks, which would be undetectable on the short timescales available so far.

\section{Acknowledgments}
     We would like to thank D. C. Backer, L. A. Breakiron,
R. T. Clarke, H. Chadsey, J. DeYoung,
T. M. Eubanks, F. J. Josties, V. M. Kaspi, W. J. Klepczynski,
S. Lundgren, D.J. Nice, P.K. Seidelmann, F. Vannicola, and
G. M. R. Winkler for many helpful discussions.

\begin{table}[h]
\caption{Simple Noise Model}
\label{tab:parms}
\begin{center}
\begin{tabular}{llll}
\hline
       & White          &   White                & Random Walk        \\
       & Phase Noise    &   Frequency Noise      & Frequency Noise    \\
       & nanosecond     & $\rm 10^{-15}s/s (t/days)^{-0.5}$        &
 $\rm 10^{-15}s/s (t/days)^{+0.5}$      \\
\hline
cesium * & 0         & 25   & 0.5 \\
maser  * & 0         & 2   & 1.0  \\
PSR B1937+21 **& 100  & 0  & 3   \\
\hline
\end{tabular}                                                            \\
\end{center}

* It is assumed that the times of the phase, rate, and frequency drift
discontinuities
have been identified, and their effects removed by least-squares methods.
Such methods are required on monthly scales for masers, and much less
frequently for cesiums.   It has the effect of masking higher order variations.

** Although today's best Arecibo data for PSR B1937+21 have formal TOA errors
of 200 nanoseconds, higher precision will be achievable with the GBT
and upgraded Arecibo.  The random walk noise, computed from Monte Carlo
simulations, is numerically larger than modeled by Petit \etal \cite{ptt92}

\end{table}

\begin{table}[h]
\caption{Antenna Parameters at 1.4 GHz}
\label{tab:parms}
\begin{center}
\begin{tabular}{llllll}
\hline
                 &$T_{sys}$& Gain&Bandwidth &SNR& $T_{int}$*\\
                 &    K    & K/Jy& MHz      &2 Polarizations&Hours   \\
\hline
Arecibo, upgraded &     20   & 11  &  200     & 150&0.3\\
Arecibo, current  &    40   & 8   &    40     & 24&10 \\
GBT              &     10    & 1.9 &  200     & 50&2  \\
Effelsberg       &     20    & 1.5 &  200     & 20&14  \\
Nancay           &   40    & 1.5 &  200     & 10&55  \\
Parkes           &   20    & .64 &  320     & 11 &50 \\
Jodrell Bank     &   20    & .64 &  320     & 11&50  \\
NRAO 140$^{\prime}$&   20    & .26 &  120     & 3 &800  \\
\hline

\end{tabular}                                                            \\
\end{center}

\noindent * Time required to observe PSR J1713+0747 with a TOA error of 100
nanoseconds assuming a flux of 3 mJy and a pulse width of 18 degrees.\\
\end{table}

\begin{table}[h]
\caption{Current Pulsar Astrometric and Timing Precision}
\label{tab:parms}
\begin{center}
\begin{tabular}{lllll}
\hline
Pulsar & Observation &  Timing    & RA        & DEC \\
Name   & Duration    &  Precision & Precision & Precision  \\
       & (years)     &  $(\rm \mu sec)$ & (mas)     & (mas)  \\
\hline
PSR B1257+12\cite{wol94}   & 2.6  & 2.3 & 0.4  & 1.0  \\
PSR J1713+0747\cite{cfw94} & 1.8  & 0.4 & 0.2  & 0.3  \\
PSR B1855+09\cite{ktr94}   & 6.9  & 1.0 & 0.07 & 0.12 \\
PSR B1937+21\cite{ktr94}  & 8.2  & 0.2 & 0.03 & 0.06 \\
PSR J2019+2425\cite{nt95}  & 2.7  & 3.0 & 0.6  & 0.9  \\
PSR J2322+2057\cite{nt95}  & 2.3  & 2.9 & 1.0  & 2.0  \\
\hline
\end{tabular}                                                            \\
\end{center}

\end{table}

\clearpage

\noindent{\large\bf Figure Captions}

\begin{figure}[h]
\caption{Observed frequency stability of USNO standards as function of
averaging time.  Plotted are the average Allan deviation (rms of differences
between adjacent averages, divided by root2) of the frequencies of
6 cavity-tuned masers and 12 model 5071 cesiums as a function of
averaging time $\tau$.
The error bars indicate the ensemble rms.}
\end{figure}

\begin{figure}[h]
\caption{Frequency difference between free-running timescales of TAI,
PTB, and USNO (A.1) in fs/s.  The USNO and PTB timescales are independent,
but TAI is gently steered towards the PTB, whereas the USNO clocks'
contribution to TAI has increased to 40\%.  Nevertheless, improvement
in the last few years is obvious.  The lowest plot shows the difference
between TT and TAI, where TT represents TAI recomputed using hindsight
corrections to clock weights, offsets, and drifts.  A constant frequency
offset has been removed from all plots.  Data were obtain from the BIPM
by anonymous ftp to address 145.238.2.2
The time range shown is from
January 1982 to December 1993.
}
\end{figure}

\begin{figure}[h]
\caption  {As in previous figure, except that the vertical axis shows
the timing difference, with a
quadratic term removed.  Pulsars cannot contribute to the quadratic
terms as their true periods and spin-down rates cannot be determined.}
\end{figure}

\begin{figure}[h]
\caption {As in previous figure, except
the time range of data is shortened to include only the latest and
best data, from 1990 to 1994.
}
\end{figure}

\begin{figure}[h]
\caption{Timing residuals to pulsar parameter solutions for
PSR B1937+21 and PSR B1855+09,
from October 1984 to December 1992.
}
\end{figure}

\begin{figure}[h]
\caption{Difference between timing residuals of PSR B1937+21.  Solutions
differ only in their reference timescale, and each plot is the difference
between the solutions with the indicated timescale and with TT.
The data are from October 1984 to December 1992
}
\end{figure}

\begin{figure}[h]
\caption {Double-difference plots subtracting the TEMPO residual
differences of previous figure from the terrestrial timescale
differences.  Since pulsars cannot determine
variations below second order, a quadratic term was removed
from the terrestrial differences.
The data are from October 1984 to December 1992.
}
\end{figure}

\begin{figure}[h]
\caption{Upper part shows the residuals for a fit to pulsar parameters for
PSR B1937+21.  Lower part shows the same residuals after using a
Fourier analysis to extract all periodicities
shorter than 400 days.  No corrections have been made for spectral
leakage or aliasing.  The data are from October 1984 to December 1992.
}
\end{figure}

\begin{figure}[h]
\caption{Transfer functions for 165 observations equally
spaced over 8 years, and for 330 observations equally spaced over
16 years as a function of period (1/frequency).  In this plot,
a value of 1 indicates no noise is
absorbed in the fit to pulsar parameters, and a value of 0 indicates
that all noise is absorbed.
}
\end{figure}


\bigskip

\begin{thebibliography}{9}

\bibitem{b82}
Backer, D.C., Kulkarni,S.R.,Heiles,C.E.,Davis,M.M., Goss, W.M. (1982)
\newblock {\it Nature}, {300},615.

\bibitem{ktr94}
Kaspi, V.M., Taylor, J.H., Ryba, M.F. (1994)
\newblock {\it Astrophys. J.}, {428}, 713.

\bibitem{t94}
Thomas, C. (1995)
\newblock {\it Proc. 26th Annual PTTI}, in press.

\bibitem{fb90}
Foster, R.S., Backer, D.C. (1990)
\newblock {\it Astrophys. J.} {361}, 300.

\bibitem{gp91}
Guinot, B., Petit, G. (1991)
\newblock {\it Astron. Astrophys.} {248}, 292.

\bibitem{ptt93}
Petit, G., Thomas, C., Tavella, P. (1993)
\newblock {\it Proc. 24th Annual PTTI}, 73.

\bibitem{m71}
Mulholland, J.D. (1971)
\newblock {\it Astrophys. J.} {165}, 105.

\bibitem{b95}
Backer, D.C. (1995)
\newblock {\it Proc. Marcel Grossmann Meeting}, in press

\bibitem{srtr90}
Stinebring, D.R., Ryba, M.F., Taylor, J.H.,  Romani, R.W. (1990)
\newblock {\it Phys.\,Rev.\,Lett.}, { 65}, 285.

\bibitem{fc90}
Foster, R.S., Cordes, J.M. (1990)
\newblock {\it Astrophys. J.} {364}, 123.

\bibitem{lrwn90}
Lestrade, J.F., Rogers, A.E.E., Whitney, A.R., Niell, A.E.,
Phillips, R.B., Preston, R.A. (1990)
\newblock {\it Astron. J.} {99}, 1663.

\bibitem{gl92}
Ghosh P., Lamb, F.K. (1992)
\newblock In: {\it X-ray Binaries and Formation of Binary and Millisecond
Pulsars},
eds. E.P.J. van den Heuvel \& S.A. Rappaport, Dordrecht: Kluwer, 487

\bibitem{bl95}
Bailes, M., Lorimer, D.R. (1995)
\newblock In: {\it Millisecond Pulsars: A Decade of Surprise}, ed Fruchter,
Tavani and Backer, 17.

\bibitem{fcwa95}
Foster, R. S., Cadwell, B. J., Wolszczan, A., and Anderson, S. B. (1995)
\newblock {\it Astrophys. J.} {454} 1 Dec, in press.

\bibitem{tgcwm67}
Thirteenth General Conference of Weights and Measures (1967)

\bibitem{r83}
Ramsey, N.F. (1983)
\newblock {\it J.Res.}, N.B.S. {88}, 301.

\bibitem{r63}
Ramsey, N.F.,
\newblock {\it Molecular Beams} (1963)

\bibitem{bauch87a}
Bauch, A, Dorenwendt, K., Fischer, B., Heindorff, T., Muller, E.K.,
Schroder, R. (1987)
\newblock {\it IEEE Trans. Instrum. and Meas.} IM-36 {No. 2}, 613.

\bibitem{cg92}
Cutler, L.S., Giffard, R.P. (1992)
\newblock {\it Proc. 1992 IEEE Frequency Control Symposium}, 127.

\bibitem{br95}
Breakiron, L. (1995)
\newblock {\it Proc. 26th Annual PTTI}, in press.

\bibitem{w94}
Wheeler, W.J., Chaulmers, D.N., McKinley, A.D., Kubik, A.J., Powell, W. (1995)
\newblock {\it Proc. 26th Annual PTTI}, in press.

\bibitem{bauch87b}
Bauch, A, Dorenwendt, K., Heindorff, T. (1987)
\newblock {\it Metrologia} {24}, 199.

\bibitem{t51}
Townes, C.H. (1951)
\newblock {\it J. Appl. Phys.}, {22 }, 1365.

\bibitem{p73}
Peters, H.E. (1974)
\newblock {\it Proc. Fifth Annual PTTI}, 283.

\bibitem{okmp92}
Owings, H.B, Koppang, P.A., MacMillan, C.C., Peters, H.E. (1992)
\newblock {\it Proc. 1992 IEEE Frequency Control Symposium}, 92.

\bibitem{barnes71}
Barnes, J.A., Andrew, R.C., Cutler, L.S., Healey, D.J., Leeson, D.B.,
McGunigal, T.E., Mullen, J.A.,Smith, W.L., Sydnor, R.L., Vessot, R.F.C. ,
Winkler, G.M.R. (1971)
\newblock {\it IEEE Trans. Instr. and Meas.} IM-20 {2}, 105.

\bibitem{mkdgc95}
Matsakis, D., Kubik, A.T., DeYoung, J., Giffard, R, Cutler, L. (1995)
\newblock {\it Proc. 1993 IEEE Frequency Control Symposium}, in press.

\bibitem{tpdm94}
Tjoelker, R.L., Prestage, J.D., Dick, G.J., Maleki, L. (1994)
\newblock {\it Proc. 1994 IEEE Frequency Control Symposium}, 739.

\bibitem{fslcmb94}
Fisk, P.T.H., Sellars, M.J., Lawn, M.A., Coles,.C, Mann, A.G.,
Blair, D.G. (1994)
\newblock {\it Proc. 1994 IEE Frequency Symposium}, 731.

\bibitem{whw91}
Wineland, D.J., Heinzen, D.J., Weimer, C.S. (1991)
\newblock {\it Proc. 22nd Annual PTTI}, 53.

\bibitem{kprsj95}
Kastberg, A., Phillips, W.D, Rolston, S.L., Spreeuw, R.J.C, Jessen, P.S.
 (1995)
\newblock {\it Phys. Rev. Letters} {74}, 1542.

\bibitem{d91}
Dixon, T.H (1991)
\newblock {\it Reviews of Geophysics}, {29}, 249.

\bibitem{k83}
Klepczynski, W.J. (1983)
\newblock {Proc. IEEE} {71}, No. 10, 1193.

\bibitem{lt91}
Lewandowski, W., Thomas, C. (1991)
\newblock {\it Proc. IEEE} {79}, No. 7, 991.

\bibitem{c93}
Chadsey, H. (1994)
\newblock {\it Proc. 25th Annual PTTI}, 317.

\bibitem{dey94}
DeYoung, J.A., Klepczynski, W.J., McKinley, A.D., Powell, W., Hetzel, P.,
Bauch, A., Davis, J.A.,Pearce, P.R., Baumont, F, Claudon,P, Grudler, P.,
de Yong, G., Kirchner, D., Ressler, H. Soring, A., Hackman, C, Veenstra, L.
(1995)
\newblock {\it Proc. 26th Annual PTTI}, in press.

\bibitem{bipm93}
BIPM
\newblock {\it Annual Report} (1993)

\bibitem{dlgs93}
Drullinger, R.E., Lowe, J.P., Glaze, D.J., Shirley, J. (1993)
\newblock {\it Proc. IEEE Frequency Symposium}, 71.

\bibitem{br91}
Breakiron, L. (1992)
\newblock {\it Proc. 23rd Annual PTTI}, 297.

\bibitem{whp70}
Winkler, G.M.R., Hall, R.G., Percival, D.B. (1970)
\newblock {\it Metrologia} {6}, 126.

\bibitem{e93}
Eubanks, T.M. (1993)
\newblock {\it Contributions of Space Geodesy to Geodynamics: Earth Dynamics}.
\newblock {\it Geodynamics} {24}, 1.

\bibitem{m72}
Mulholland, J.D. (1972)
\newblock {\it PASP} {84}, 357.

\bibitem{a82}
Aoki, S., Guinot, B., Kaplan, G.H., Kinoshita, H.,
McCarthy, D.D., Seidelmann, P.K. (1982)
\newblock {\it Astron. Astrophys.} {105}, 359.

\bibitem{neos94}
U.S. National Earth Orientation Service (1995)
\newblock {\it Annual Report}

\bibitem{g87}
Guinot, B. (1987)
\newblock {\it Metrologia} {24}, 195.

\bibitem{iau91}
IAU Resolution A4
\newblock {\it Proc. 21st General Assembly} Transactions of the IAU, XXI.
Kluwer, Dordrecht (1992)

\bibitem{sf92}
Seidelmann, P.K, Fukushima, T. (1992)
\newblock {\it Astron. Astrophys.} {265}, 833.

\bibitem{f95}
Fukushima, T. (1995)
\newblock {\it Astron. Astrophys.} {294}, 895.

\bibitem{b94}
Backer, D.C. (1994)
\newblock {\it IAU Symposium} {165}.

\bibitem{b91}
Backer, D.C. (1991)
\newblock {\it Springer-Verlag Lecture Notes in Physics} {418}, 193.

\bibitem{fip86}
Friedman, J.L., Ipser, J.R., Parker, L., (1986)
\newblock {\it Astrophys. J.}, {304}, 115.

\bibitem{ctktr94}
Camilo, F., Thorsett, S.E., Kulkarni, S.R. (1994)
\newblock {\it Astrophys. J.}, {421}, L15.

\bibitem{bra87}
Blandford, R.D., Romani,R.W., Applegate, J.H. (1987)
\newblock {\it Mon. Not. R. Astron. Soc.}{225},51P.

\bibitem{wol89}
Wolszczan, A., Kulkarni, S.R., Middleditch, J., Backer, D.C.,
Fruchter, A.S., Dewey, R.J. (1989)
\newblock {\it Nature}, {337}, 531.

\bibitem{s90}
Standish, E.M. (1990)
\newblock {\it Astron. Astrophys.}, { 233}, 252.

\bibitem{c95}
Taylor, J.H., Manchester, D.N.,  Lyne, A.G., and Camilo, F (1995)
\newblock {\it in prep, anonymous ftp}, [pulsar.princeton.edu].

\bibitem{dd86}
Damour, T., Deruelle, N. (1986)
\newblock {\it Ann. Inst. H.Poincare (Physique Theorique)} {44}, 263.

\bibitem{wol94}
Wolszczan, A. (1994)
\newblock {\it Science}, {264}, 538.

\bibitem{tw89}
Taylor, J.H., Weisberg, J.M. (1989)
\newblock {\it Astrophys. J.}, { 345}, 434.

\bibitem{cblb94}
Cognard,I., Bourgois, G., Lestrade, J.F., Biraud, F. (1995)
\newblock In: {\it Millisecond Pulsars: A Decade of Surprise}, ed Fruchter,
Tavani and Backer, 372.

\bibitem{antt94}
Arzoumanian, Z., Nice, D.J., Taylor, J.H.,  Thorsett, S.E. (1994)
\newblock {\it Astrophys. J.}, { 422}, 671.

\bibitem{fcw95}
Foster, R.S., Camilo, F., Wolszczan, A. (1995)
\newblock {\it Proc. Marcell Grossmann Meeting}, in press

\bibitem{bnr84}
Blandford, R., Narayan, R., Romani, R.W. (1984)
\newblock {\it J. Astrophys. Astr.} {5}, 369.

\bibitem{k94}
Kaspi, V.M. (1994)
\newblock {\it Ph.D. Thesis}, Princeton University.

\bibitem{ptt92}
Petit, G., Tavella, P., Thomas, C. (1992)
\newblock {\it Proc. 6th European Frequency and Time Forum}, 57.

\bibitem{cfw94}
Camilo, F., Foster, R.S., Wolszczan, A. (1994)
\newblock {\it Astrophys. J.} {437}, L39.

\bibitem{nt95}
Nice, D.J., Taylor, J.H. (1995)
\newblock {\it Astrophys. J.} {441}, 429.


\end{thebibliography}
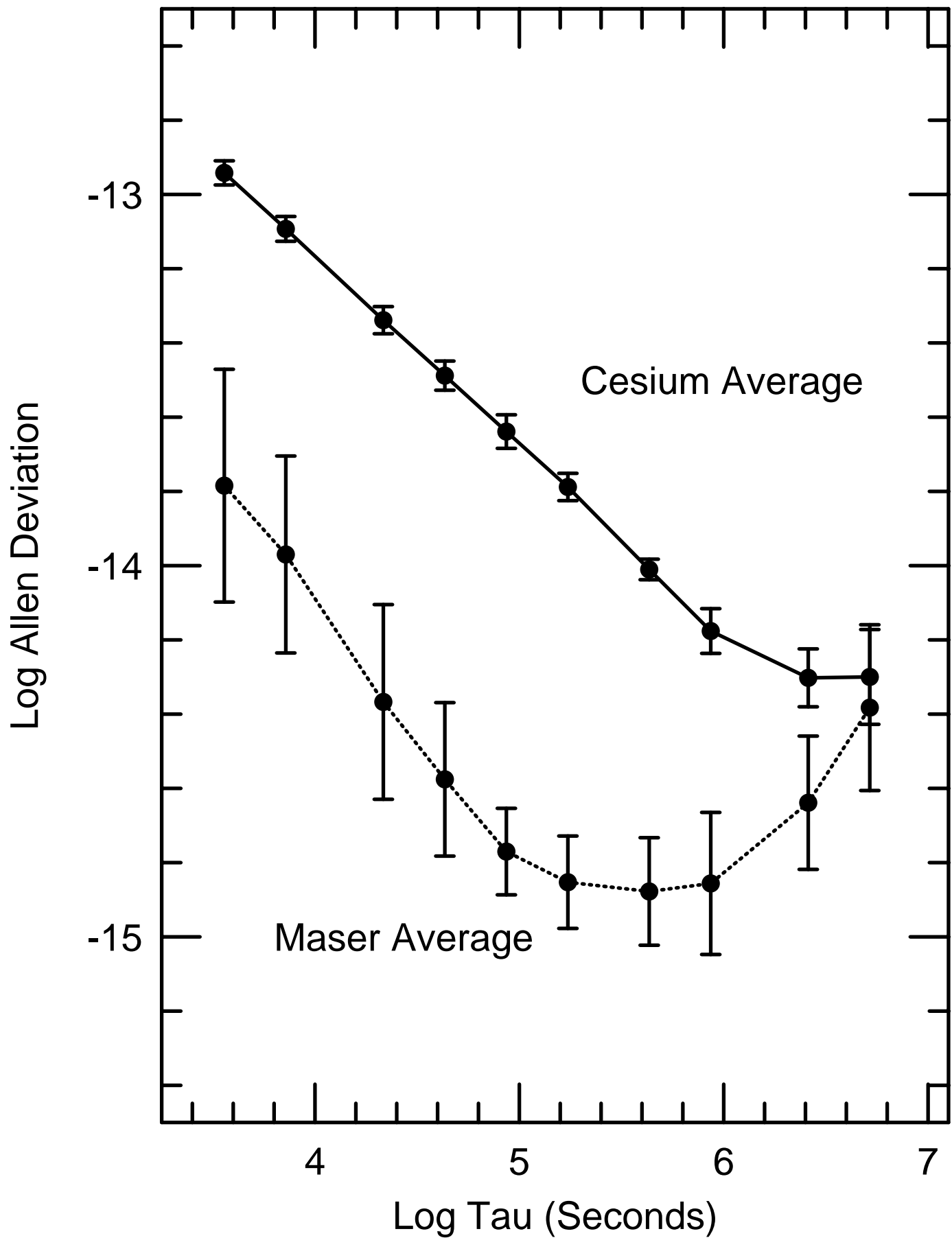
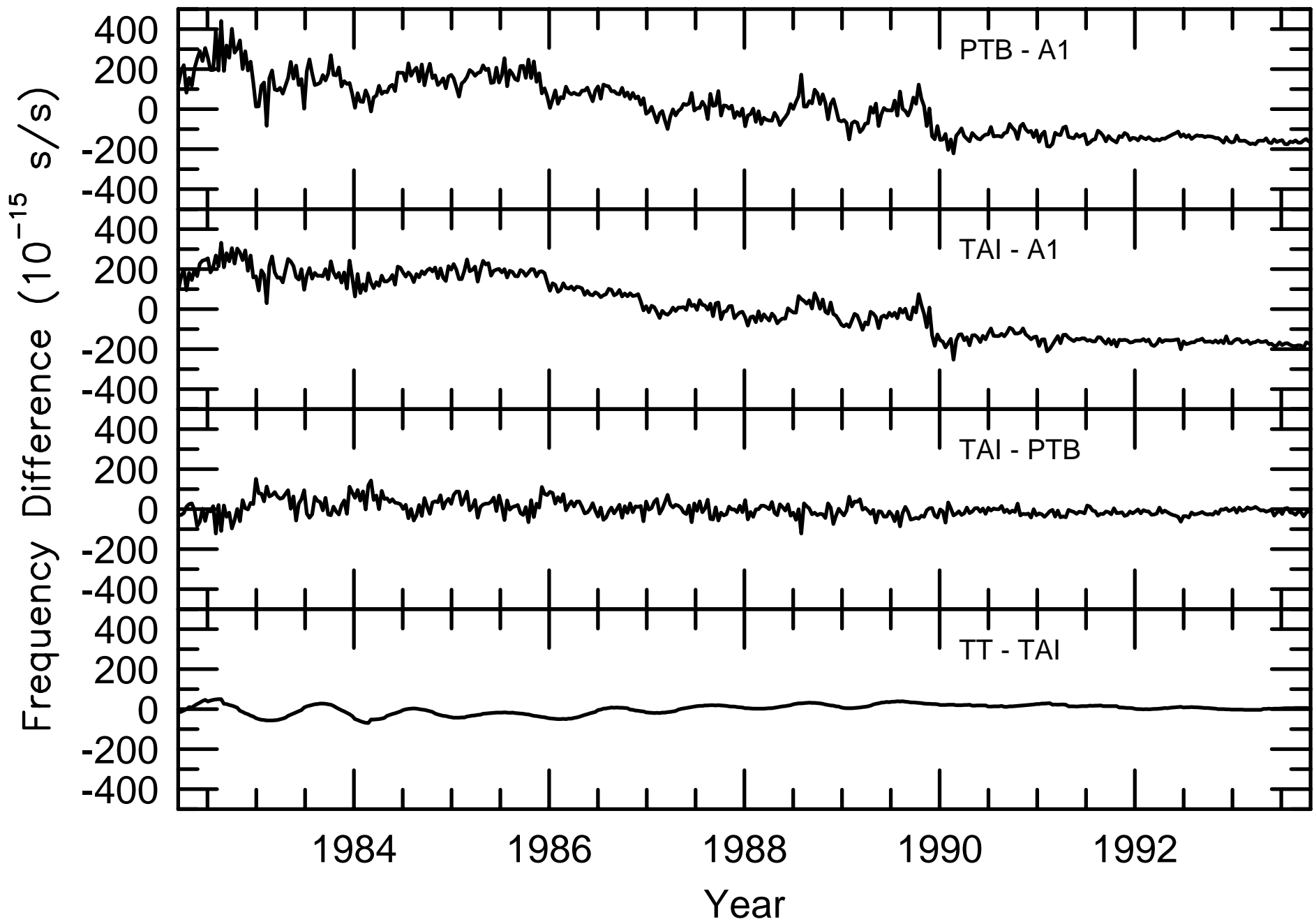
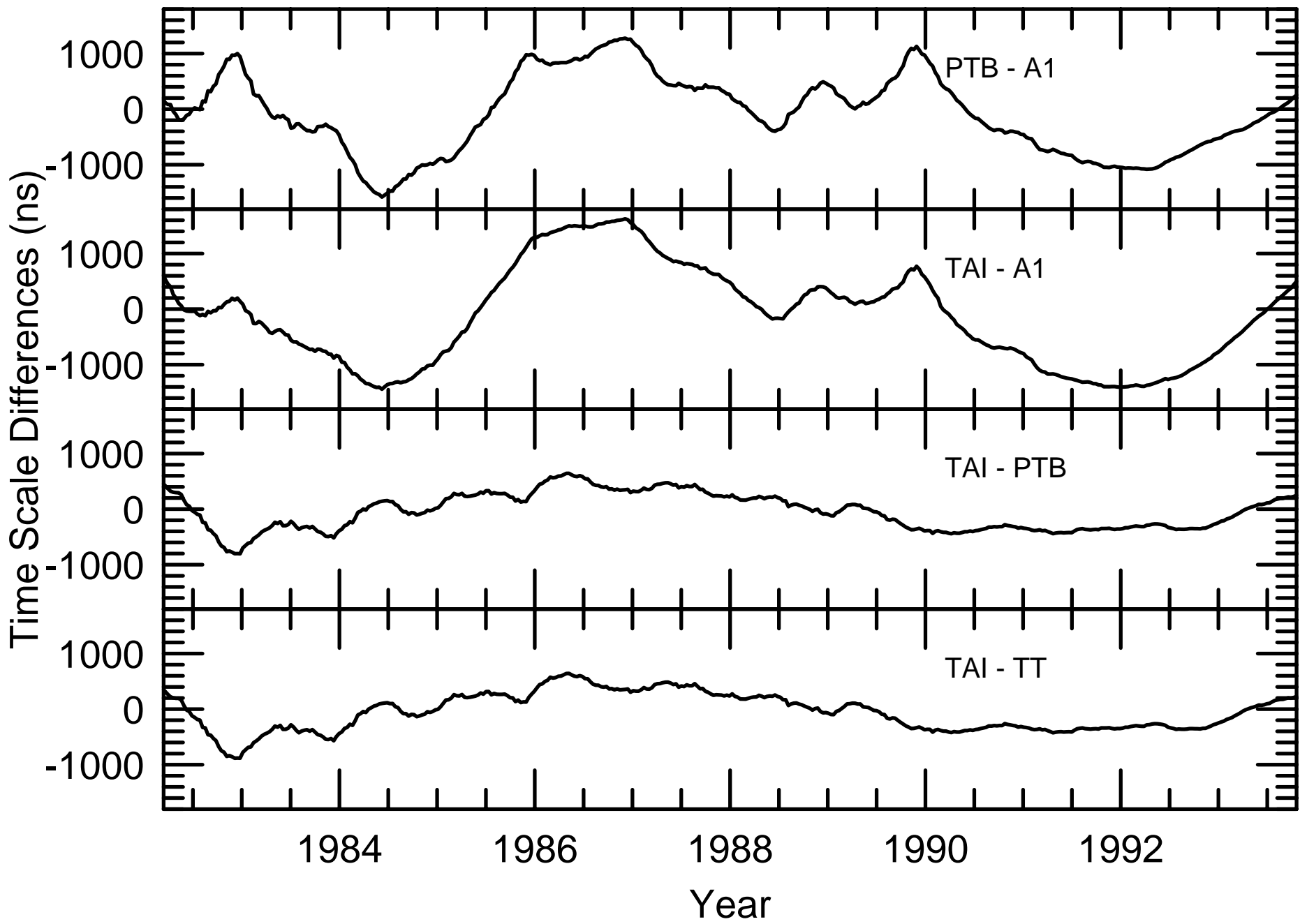
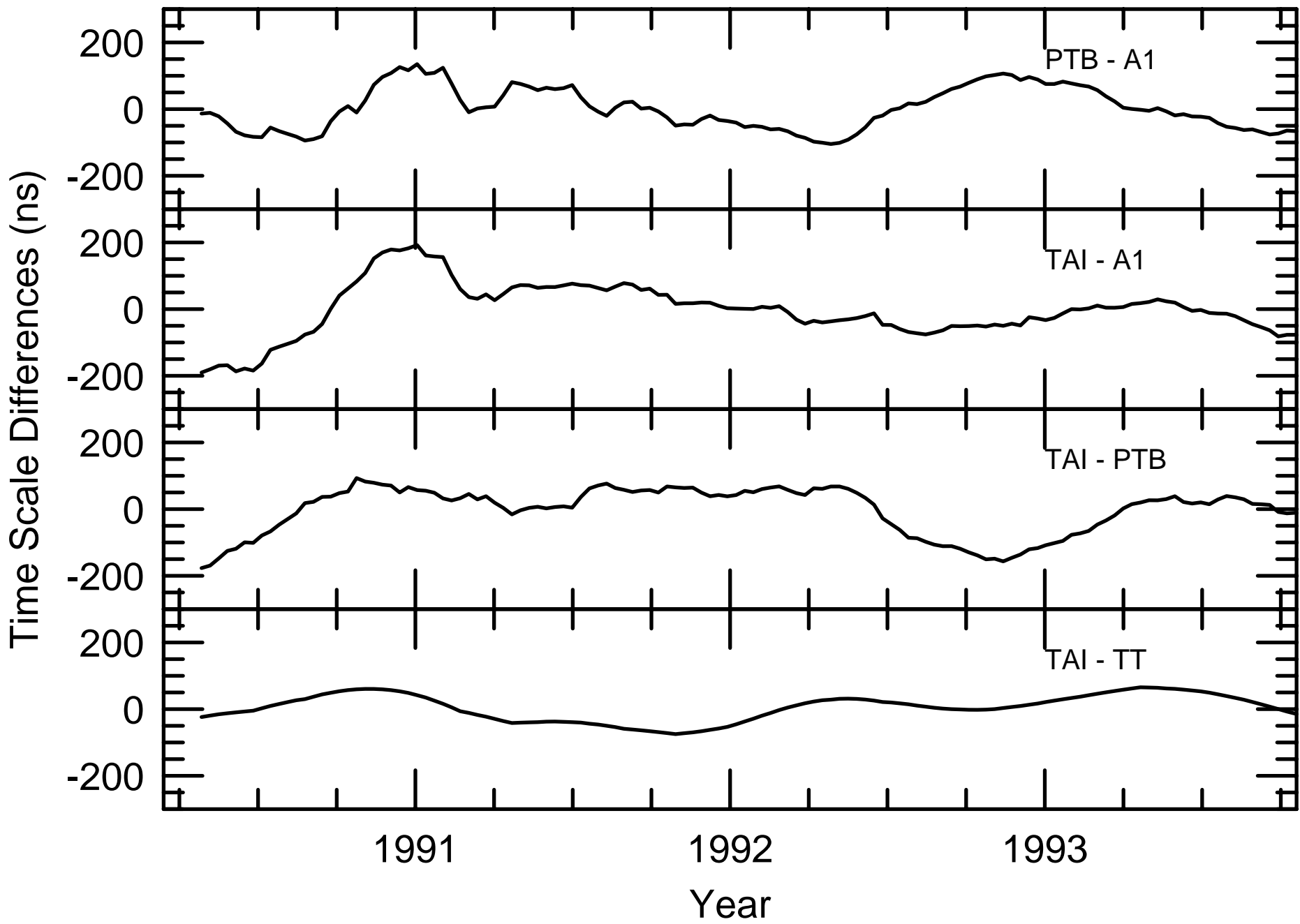
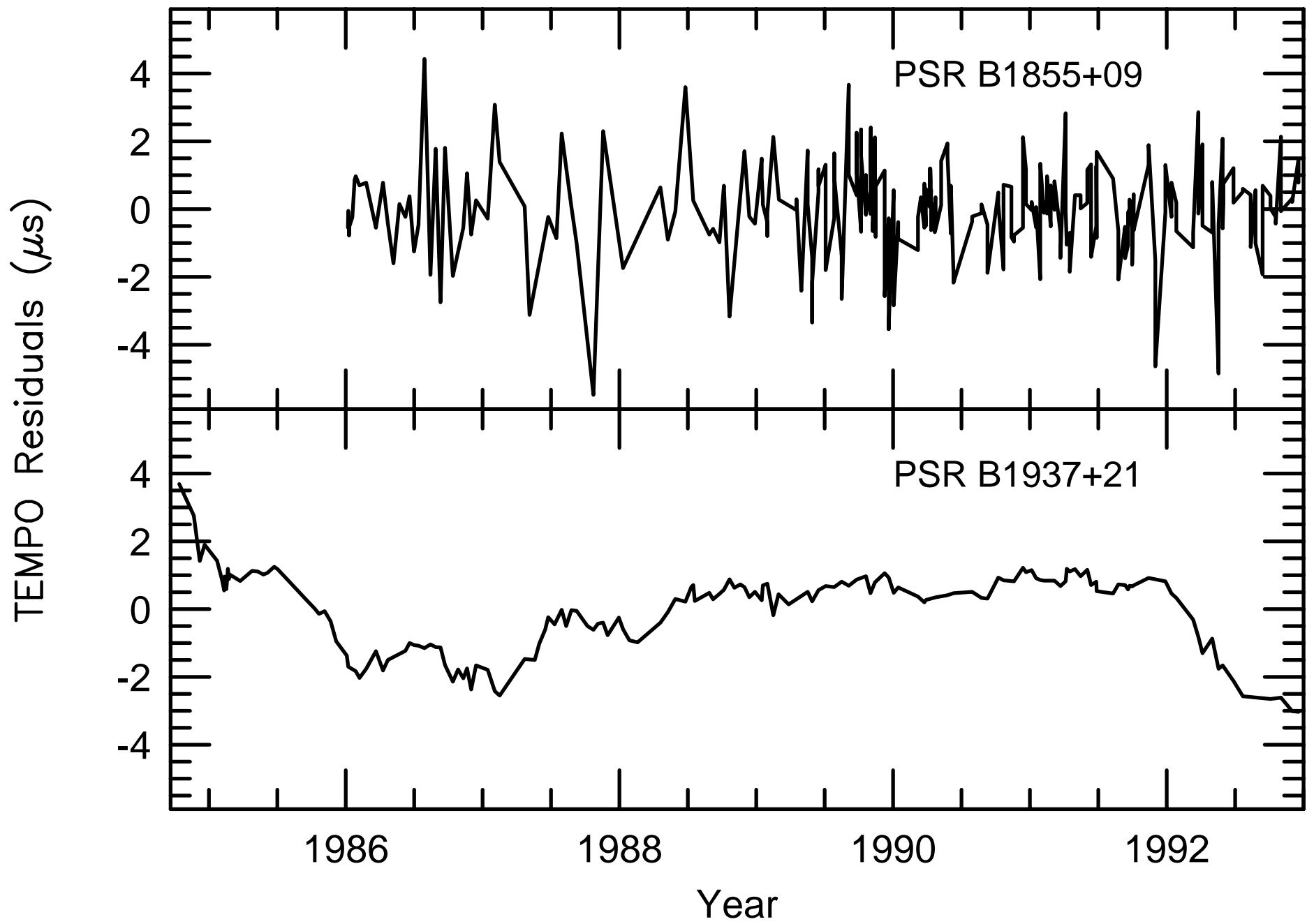
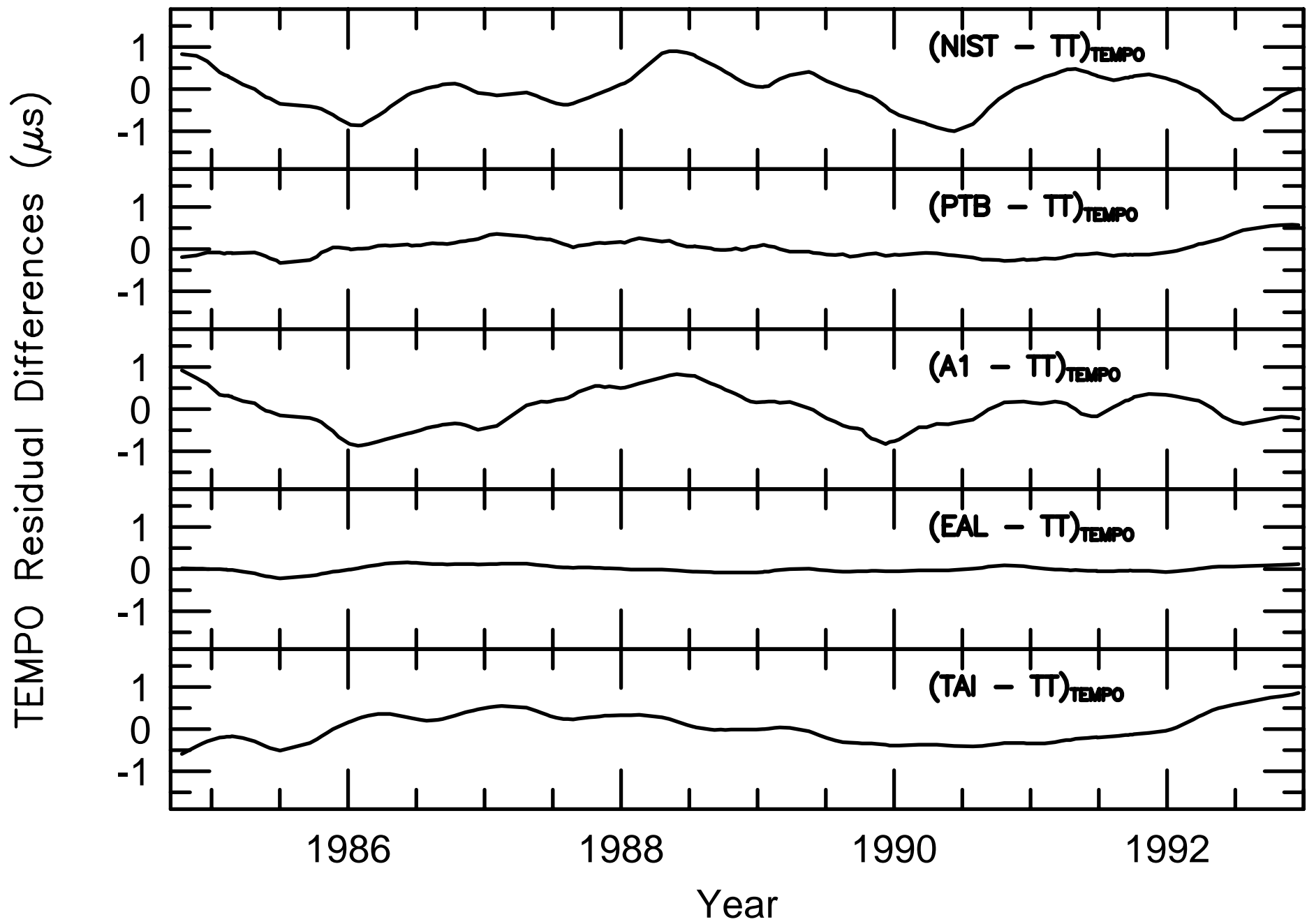
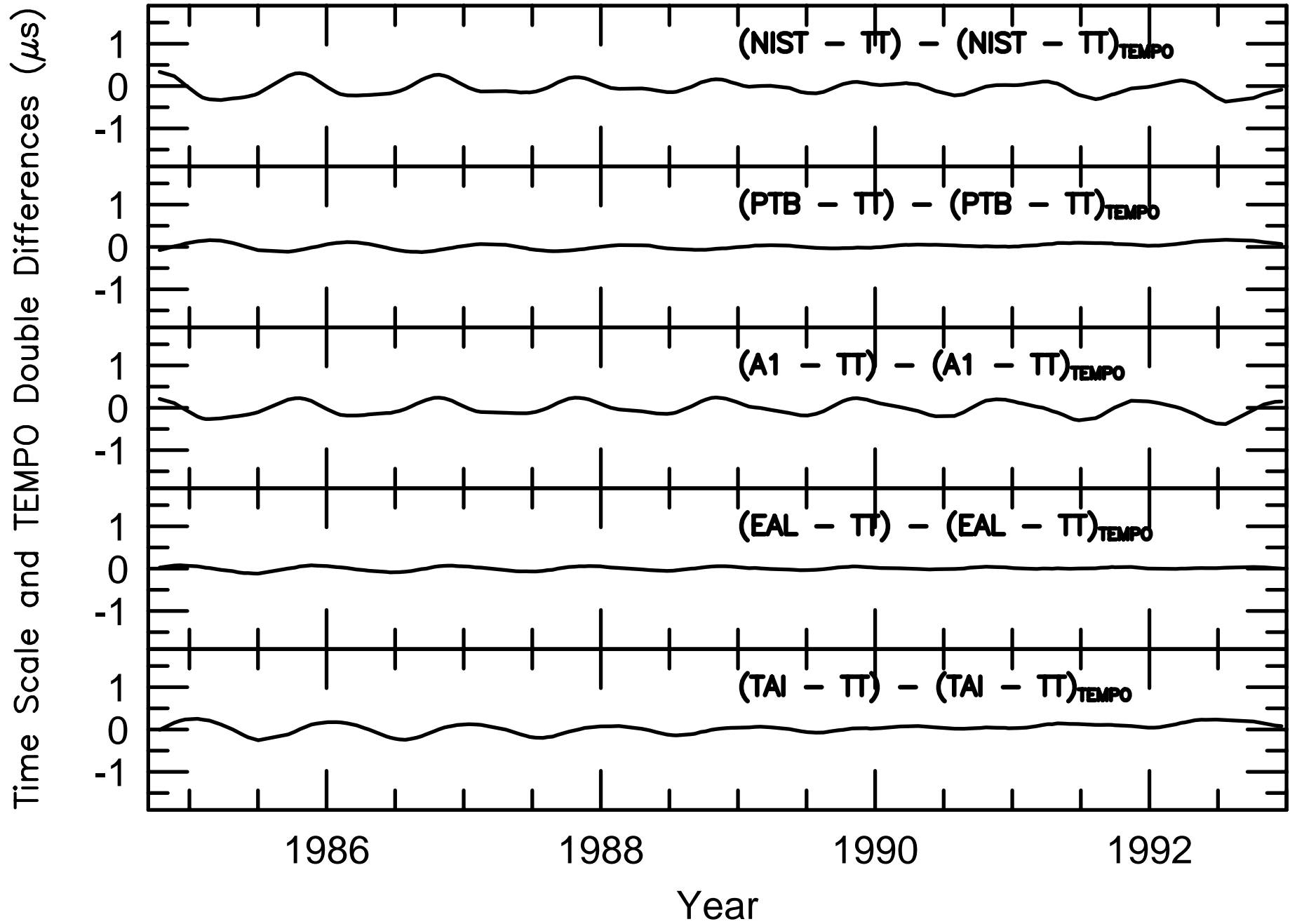
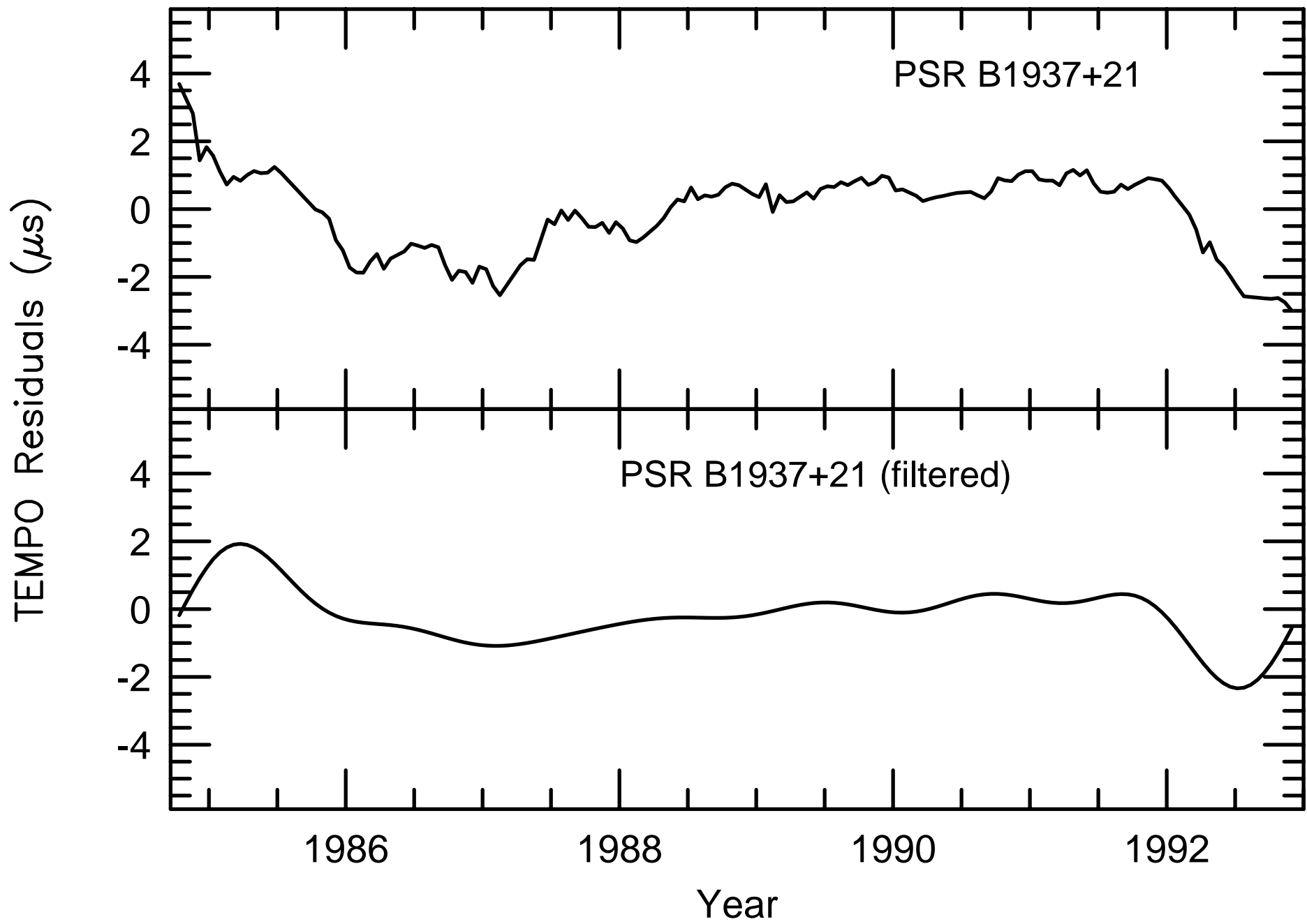
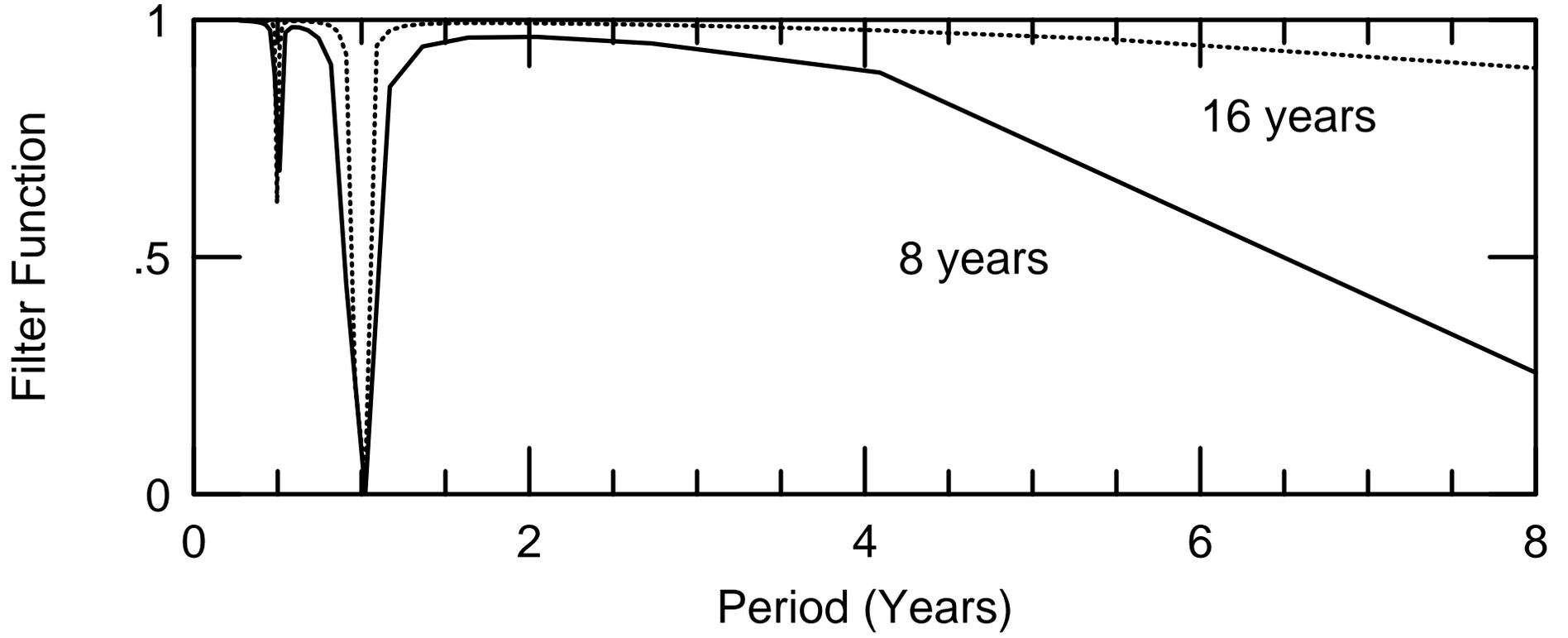
\end{document}